\documentclass[aps,prl,twocolumn,showpacs,amsmath,amssymb]{revtex4}
\usepackage{graphicx}
\usepackage{amsmath}
\usepackage{amsbsy}
\usepackage{dcolumn}
\usepackage{bm}

\begin{document}

\title{Stable two-channel Kondo fixed point of an SU(3)
quantum defect in a metal:\newline renormalization group 
analysis and conductance spikes}

\date{\today}
\author{Michael Arnold, Tobias Langenbruch and Johann Kroha}
\affiliation{Physikalisches Institut, Universit\"at Bonn,
Nussallee 12, 53115 Bonn, Germany}

\begin{abstract}
We propose a physical realization of the two-channel Kondo (2CK) 
effect, where a dynamical defect in a metal has a unique ground state 
and twofold degenerate excited states. 
In a wide range of parameters the interactions with the electrons 
renormalize the excited doublet downward below the bare defect ground 
state, thus stabilizing the 2CK fixed point. 
In addition to the Kondo temperature $T_K$ this system
exhibits another low-energy scale, associated
with ground-to-excited-state transitions, which can be exponentially
smaller than $T_K$. Using the perturbative nonequilibrium 
renormalization group we demonstrate that this can 
provide the long-sought explanation of the sharp conductance spikes
observed by Ralph and Buhrman in ultrasmall metallic point contacts.
\end{abstract}
\bigskip
\noindent 
\maketitle 

The two-channel Kondo (2CK) effect has been intriguing physicists ever
since it was proposed by Nozi\'eres and Blandin in 1980 \cite{nozieres80}
because of its exotic non-Fermi liquid ground state with a nonvanishing
zero-point entropy $S(0)=k_B \ln \sqrt{2}$ \cite{andrei84}. 
It arises when a discrete, degenerate quantum degree of freedom is 
exchange-coupled to two conserved conduction electron continua or 
channels in a symmetrical way \cite{coxzawa98}. 
Early on, Vl\'{a}dar and Zawadowski \cite{vladar82} put forward two-level 
systems (TLS), e.g. an atom in a double well potential embedded in a 
metal, as a physical realization of a 2CK system. 
However, this model encountered 
difficulties because for small exchange coupling, $J\ll 1/N(0)$, 
with $N(0)$ the conduction density of states
at the Fermi energy, the renormalized level splitting 
of the TLS cannot become less than the exponentially small Kondo
temperature, $T_K \sim \exp [-1/(4N(0)J)]$, and a large effective 
coupling, $N(0)J\approx 1$, cannot be realized even by
invoking enhanced electron assisted tunneling via virtually excited 
states of the TLS \cite{zawa94} because of cancellation 
\cite{aleiner01} and screening \cite{aleiner02} effects. 
It remains difficult to reach by a resonance 
enhancement of $N(0)$ proposed in Ref.~[\onlinecite{zarand05}].

Signatures of the 2CK effect have been observed experimentally in 
heavy fermion compounds \cite{cichorek05} and most recently in 
especially designed quantum dot systems \cite{goldhaber07}.
However, perhaps the most intense interest has been provoked by the 
conductance anomalies observed by Ralph and Buhrman \cite{ralph92} 
in ultrasmall metallic point contacts. This is because, on the one
hand, these measurements showed most clear-cut 2CK signatures near  
zero bias, including the theoretically expected 
\cite{affleck91,hettler94} scaling behavior \cite{vdelft95}, 
and, on the other hand, these zero-bias anomalies (ZBA) occurred in 
seemingly simple Cu point contacts. In addition, sharp spikes were 
observed in the conductance at elevated bias \cite{ralph92}, 
such that contacts with conductance spikes always exhibited 
also the ZBA.  
These results have remained essentially not understood to the 
present day mainly due to the lack of a microscopic model with a
stable 2CK fixed point, although some of the features could alternatively 
be accounted for by a distribution of TLS without invoking the 2CK 
effect \cite{kozub97}. However, a reliable understanding requires
an explanation of all the experimental features.
The difficulty resides in the fact that in the usual 2CK models 
the Kondo SU(2) symmetry is easily destroyed by relevant perturbations.

\begin{figure}[t]
\begin{center}
\includegraphics[width=0.9\linewidth]{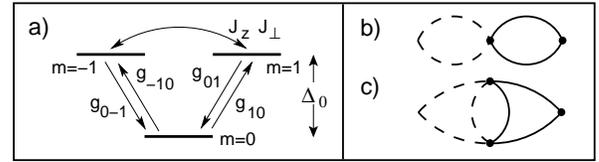}
\end{center}
\vspace*{-1.2em}
\caption{\label{fig1}
a) Level scheme of the three-state defect, defining the 
coupling constants in Eq.~(\ref{hamiltonian});  
b) 2nd and c) 3rd order contributions to the 
current through a point contact with Kondo impurities. ---~: 
conduction electron, - - -~: defect state propagators. 
A $\bullet$ indicates a sum over all 4-point vertices 
of Eq.~(\ref{hamiltonian}).  
}
\vspace*{-0.0cm}
\end{figure}

In this Letter we present a realistic, microscopic model for the 2CK effect,
where the Kondo degree of freedom are the parity-degenerate rotational
states ($m=\pm 1$) of an atomic three-level system (3LS)  
and the channel degree of freedom is the magnetic conduction 
electron spin. Such a level scheme (see Fig.~\ref{fig1}) is generically 
realized by an atom (hydrogen ion) moving in a modulated Mexican 
hat potential which, e.g., is formed in the interstitial space of 
the (111) plane of a Cu lattice.  
Although similar models have been considered 
before \cite{moustakas96,zarand96}, they have not been analyzed in
detail, especially not in a current-carrying, finite bias situation
\cite{rosch03,paaske04a,paaske04b}. 
We show that for a wide range of parameters the 2CK fixed point is 
stabilized by a correlation-induced level crossing, forcing the 
degenerate defect levels to be the interacting ground states of the defect. 
In addition to the 2CK 
Kondo temperature $T_K$, the 3LS has another correlation-induced
low-energy scale $T_K^{\star}$, associated with
ground-to-excited-state transitions. $T_K^{\star}$ can be exponentially 
smaller than $T_K$. We show by explicit renormalization group (RG)
calculations of the non-equilibrium differential conductance 
\cite{rosch03,paaske04a} that this can account for conductance
spikes at elevated bias, much narrower than the ZBA.

{\it The model. ---}
For the defect levels we employ the Abrikosov pseudospin 
representation, where $f^{\dagger}_{m}$ is the creation operator for the
defect in state $m=0,\pm 1$, $m=0$ labeling the (non-interacting)
ground state and $m=\pm 1$ the excited doublet [cf.~Fig.~\ref{fig1}a)].  
The defect dynamics are subject to the constraint 
$\hat Q = \sum_{m=0,\pm 1} f^{\dagger}_{m} f^{\phantom{\dagger}}_{m} =1$.
The defect coupled to the conduction electron continuum is then 
described by the Hamiltonian,
\begin{eqnarray}
H&=&{\sum_{{\bf k}\sigma m \alpha}}' 
    \varepsilon_{{\bf k}} c^{\alpha\ \ \dagger}_{{\bf k}\sigma m}
                         c^{\alpha\ \ \phantom{\dagger}}_{{\bf k}\sigma m}
  +\Delta_0 \sum_{m=\pm 1} f^{\dagger}_{m}
                         f^{\phantom{\dagger}}_{m}  \nonumber\\ 
  &+&\sum_{\sigma\alpha\beta}
     \left[ \frac{J_z}{2} S_z  s_z^{\sigma\ \alpha\beta}  
           +J_{\perp}\left( S_{1,-1}^{\phantom{\sigma}} s_{-1,1}^{\sigma\ \alpha\beta} 
                          + S_{-1,1}^{\phantom{\sigma}} s_{1,-1}^{\sigma\ \alpha\beta}
                     \right) 
     \right]  \nonumber \\
 &+& 
\sum_{\stackrel{{\bf k}{\bf k}'}{\sigma\alpha\beta}}
\sum_{\stackrel{m,n}{-1\leq n-m \leq 1}}
\left[
 g_{m0}^{(n)} S _{m,0}^{\phantom{\sigma}} s^{\sigma\ \alpha\beta}_{n-m,n} + 
 H.c. 
\right] \ , 
\label{hamiltonian}
\end{eqnarray}
where the first term represents the conduction electron kinetic energy, the 
second one the degenerate local doublet with level splitting $\Delta_0$
above the defect ground state, and the third and fourth terms 
transitions between the local levels induced by conduction electron
scattering, see Fig.~\ref{fig1}a). 
The conduction electron operators, $c^{\alpha\ \ \dagger}_{{\bf k}\sigma m}$, 
carry the conserved magnetic spin  $\sigma=\pm 1/2$ as well as an SU(3) index 
$m=0,\pm 1$ which is altered by the defect scattering, and which is thought
to be associated with an electronic orbital degree of freedom
centered around the defect. The prime in the kinetic energy indicates a 
restricted momentum sum such that $\sum_{{\bf k}m} ' \equiv \sum _{{\bf k}}$
covers the complete momentum space. 
Note that in the rotational defect model the effective band cut-off reduction
of Ref.~\cite{aleiner02} due to screening by high-energy electrons does not
occur, because the transitions $m=1 \leftrightarrow m=-1$ are not associated
with a charge redistribution. For later use we assume that the 
defect is located in a nanoscopic point 
contact between two metallic leads held at chemical potentials 
$\mu _{\pm}=\mu \pm V/2$ \cite{ralph92}. 
The electron distribution function at the location of the
defect is then $f(\omega)=[f_0(\omega + V/2)+f_0(\omega - V/2)]/2$,
with $f_0(\omega)$ the Fermi distribution function. 
Throughout, Greek superscripts $\alpha, \beta = \pm$ indicate 
whether an electron is from the high ($\mu _+$) 
or the low  ($\mu _-$) potential reservoir. 
The defect SU(3) operators are defined as  
$S_z=f^{\dagger}_{1} f^{\phantom{\dagger}}_{1}
    -f^{\dagger}_{-1} f^{\phantom{\dagger}}_{-1}$,
and $S_{mn}=f^{\dagger}_{m} f^{\phantom{\dagger}}_{n}$,
and the SU(3) operators acting on the electronic Fock space, 
$s_z$, $s_{mn}$, are obtained by substituting
$f^{\phantom{\dagger}}_{m} \rightarrow 
 \sum_{{\mathbf k}}c^{\alpha\ \ \phantom{\dagger}}_{{\bf k}\sigma m}$ 
in the above expressions. 
In Eq.~(\ref{hamiltonian}) we have chosen a representation of SU(3)
which explicitly exhibits the unbroken symmetry of the SU(2)
subgroup in the states $m=\pm 1$.  
In the following we use dimensionless couplings 
denoted by the identification $N(0)g_{mn}^{(j)}\rightarrow g_{mn}^{(j)}$.

\begin{figure}[b]
\begin{center}
\includegraphics[width=0.9\linewidth]{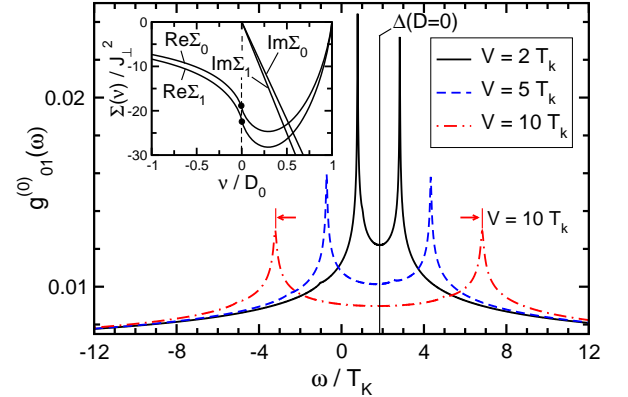}
\end{center}
\vspace*{-1.2em}
\caption{\label{fig2}
(Color online)
Coupling function $g_{01}^{(0)}(\omega)$ for different 
bias voltages $V$ (parameter values as in Fig.~\ref{fig3}). 
The peak structure of $g_{01}^{(0)}(\omega)$
is centered around the renormalized level spacing $\Delta (D = 0)$ 
and is split by the respective bias $V$. The inset shows 
the defect selfenergy $\Sigma_m(\omega)$, $m=0,1$, calculated in bare 
2nd order PT. The level spacing renormalization is the difference between the
two solid dots, 
$\delta \Delta = {\rm Re}(\Sigma _1(\Delta)-\Sigma _0(\Delta))$.
}
\end{figure}

{\it Nonequilibrium RG. ---}
The defect is described by the (retarded) impurity propagator
$G_{m}(\nu)=[\nu - |m|\Delta + i\Gamma_m ]^{-1}$, which, by symmetry, remains 
diagonal in $m$, even when coupled to the electron system. Here
$\Delta$ is the renormalized level-spacing (see below), 
and $\Gamma_m$ is the interaction-induced decay rate of level $m$.  
When a bias voltage $V$ is applied, all electrons within at least the 
voltage window contribute to the scattering.
The corrections to the electron--defect vertex depend on 
the energy of the incoming electron, $\omega$, 
and of the in- and out-going defect states $n$, $m$. 
To a very good approximation \cite{rosch03}, the defect dynamics 
can be taken on-shell ($\nu=|m|\Delta$), so that the electron--defect vertex
depends on $n$, $m$, and on $\omega$ as the only continuous variable.
This has led in Ref.~\cite{rosch03} to the concept of a perturbative RG for 
coupling {\it functions}, parameterized by the energy of the incoming electron,
i.e. the Hamiltonian, considered at each electron energy $\omega$, has
its own RG flow. The one-loop RG equations for our SU(3) model can be derived 
as \cite{rosch03}, 
\begin{widetext}
\begin{eqnarray}
\frac{d g_{mn}^{(j)\ \alpha\beta}(\omega)}{d \ln D} &=&
2 \sum _{\stackrel{j \ell  \gamma}{-1\leq j+n-\ell \leq 1}}  
   g_{m\ell}^{(j+n-\ell)\ \alpha\gamma}
   \bigl(\Omega _{n\ell} \bigr) \, 
   g_{\ell n}^{(j)\ \gamma\beta}
   \bigl(\omega \bigr) \, 
   \Theta\Bigl(D-\sqrt{\left(\Omega_{n\ell}-\frac{\gamma V}{2} \right)^2+\Gamma_{\ell}^2}\Bigr)
   \bigl(1-\delta_{m\ell}\delta_{n\ell}\bigr) - {\rm exch.}  
\label{RG}
\end{eqnarray}
\end{widetext}
with $g_{1,-1}^{(1)}=g_{-1,1}^{(-1)}=J_{\perp}$ 
and $g_{mm}^{(j)} = jm\,J_{z}/2$, for $j=\pm 1$, $m=\pm 1$.
The $\Theta$ step function in Eq.~(\ref{RG}) controls that the 
RG flow of a particular term is cut off when the band cutoff $D$,
symmetrical about the Fermi edge $\gamma V/2$ in lead $\gamma=\pm$, 
is reduced below the energy of the intermediate scattering electron 
in that lead. This energy is $\Omega_{n\ell} = \omega+(|n|-|\ell|)\Delta(D)$. 
The Kronecker-$\delta$ factor excludes non-logarithmic terms 
which do not alter the impurity state. The energy arguments of the
$g$'s on the right-hand-side (RHS) arise from energy conservation
at each vertex. The exchange terms, exch., on the RHS of Eq.~(\ref{RG}) are 
obtained from the direct ones by interchanging in the $g$'s 
in- and out-going pseudofermion indices and by interchanging 
$m\leftrightarrow n$ everywhere. The RHS also contains the decay 
rates $\Gamma _m$ \cite{rosch03} and the renormalized level spacing $\Delta$. 
Although $\Gamma_m$ and $\Delta$ contain no leading logarithmic terms,
they acquire an RG flow, since they are calculated  
in 2nd order Keldysh perturbation theory (PT) using the running couplings 
$g_{mn}^{(j)}(\omega)$ \cite{rosch03,paaske04a}. 
Specifically, $\Delta$ is obtained during the RG flow as, 
\begin{eqnarray}
\Delta = \Delta_0 + {\rm Re}\Sigma_1(\nu=\Delta)-{\rm Re}\Sigma_0(\nu=0)\ , 
\label{levelspacing}
\end{eqnarray}
We find no significant $V$--dependence of $\Delta$ for $\ln(V/D_0)\ll 1$.
Note that the renormalized $g$'s have, in general, 
different $\omega$ dependence, even if their bare values are chosen 
equal. The typical behavior of the coupling functions as solutions of 
Eq.~(\ref{RG}) for $D\to 0$ is shown in 
Fig.~\ref{fig2} for $g_{01}^{(0)}(\omega)$ as an example.  
As expected for resonant transitions from the excited to the 
impurity ground state, $g_{01}^{(0)}(\omega)$ 
has a peak structure centered around the renormalized transition energy 
$\Delta$. 
The peak is cut off by $\Gamma _{m}$ and split by the applied bias, 
reflecting the difference between the Fermi levels in the two leads. 
The Kondo temperature
$T_K$ has been determined here and throughout as the cutoff value $D$
for which the coupling reaches $J_{\perp}(\omega =\pm V/2)=1$.

\begin{figure}[b]
\begin{center}
\includegraphics[width=0.9\linewidth]{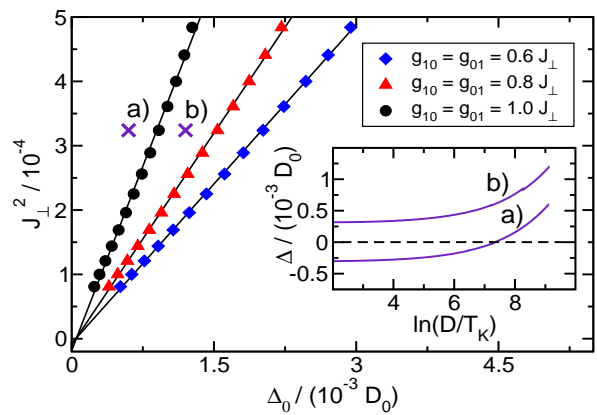}
\end{center}
\vspace*{-1.2em}
\caption{\label{fig3}
(Color online) 
Phase diagram in the $J_{\perp}^2-\Delta_0$ plane for 
$J_{z}/2= J_{\perp}=0.018$, bias $V=0$, and various values of 
$g_{0m}^{(n)}=g_{m0}^{(n)}$, $n,m=0,\pm 1$. The symbols
mark the line of instable fixed points separating 
the 2CK (left of the line) from 
the potential scattering (right of the line) 
low-temperature phase. The solid lines are straight line fits
to the $J_{\perp}^2$ vs. $\Delta_0$ plots and show that 
the level crossing is essentially an $O(J_{\perp}^2)$ PT
effect. The inset shows typical 
RG flow diagrams of the level spacing $\Delta (D)$
in a) the 2CK phase (level crossing) and b) the
potential scattering phase (no level crossing).  
}
\end{figure}

{\it Equilibrium: 2CK fixed point. ---}       
To investigate the fixed point structure of the model (\ref{hamiltonian}),
we first analyze the RG flow in equilibrium at temperature $T=0$. The 
Greek superscripts have then no relevance.
Note that even for $V=0$ the renormalized couplings have an
$\omega$-dependence. However, in this case 
it is sufficient to consider electrons
at the Fermi level only, $\omega=0$ and $\Gamma_m=0$.   
Due to time reversal symmetry we have in equilibrium, 
$g_{mn}(\omega)=g_{nm}(\omega +(|n|-|m|)\Delta)$. 
We have checked that for $V=0$ the results of the RG 
equations (\ref{RG}) coincide with those of the familiar equilibrium 
perturbative RG. As the most important point of this paper we find
that the level spacing $\Delta$ is always renormalized downward
by the RG flow and, for a wide range of bare parameters $J_{z}$, 
$J_{\perp}$, $\Delta_0$, and $g$, crosses the impurity ground level.
Such a crossing is not forbidden by symmetry, since even for 
completely symmetric initial couplings they have already 
been renormalized to different values when the level crossing occurs. 
Typical RG flow curves $\Delta (\ln D)$ are shown in the inset of 
Fig.~\ref{fig3}. It is seen that the level crossing occurs 
generically in the regime ${\rm ln}(D/T_K)\gg 1$, 
where the perturbative RG treatment is still controlled. 
Hence, this behavior can qualitatively be understood 
by 2nd order PT as follows (cf.~Fig.~\ref{fig2}, inset). 
The imaginary part of the 
defect selfenergy $\Sigma _{m}(\nu)$ is well known to have 
threshold behavior with ${\rm Im}\Sigma _{m}(\nu<0)=0$ 
\cite{kroha96}, Because of resonant transitions between the degenerate
states $m=\pm 1$ one has always 
$|{\rm Im}\Sigma _{1}(\nu)|>|{\rm Im}\Sigma _{0}(\nu)|$,
so that via Kramers-Kronig the level spacing renormalization in 
Eq.~(\ref{levelspacing}),
${\rm Re}[\Sigma _{1}(\Delta)-\Sigma _{0}(0)]$ is necessarily 
negative (the difference between the black dots 
in Fig.~\ref{fig2}, inset). We find that even in bare 2nd order 
PT for $\Sigma _{m}(\nu)$ the renormalization may be stronger 
than the bare spacing $\Delta _0$, thus inducing a level crossing.
Once the level crossing has occurred, the system is bound to flow to 
a stable 2CK fixed point, since the renormalized ground state of the
local defect is now doubly degenerate (Kondo degree of freedom), 
stabilized by spatial parity of the underlying crystal lattice, 
and the channel symmetry of the magnetic electron spin is preserved
by Kramers degeneracy. Our results are summarized in the 
equilibrium phase diagram shown in Fig.~\ref{fig3}.
The 2CK phase is realized for a sizable range of parameters.

\begin{figure}[t]
\begin{center}
\includegraphics[width=0.9\linewidth]{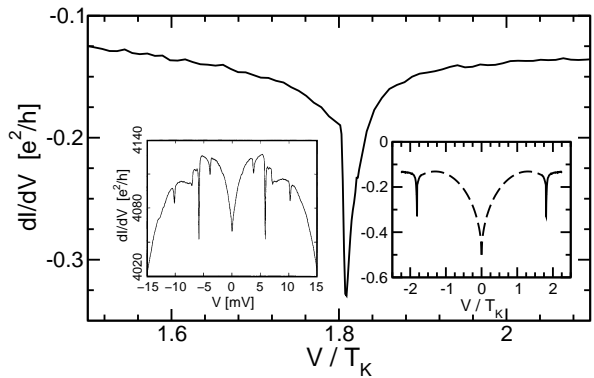}
\end{center}
\vspace*{-1.2em}
\caption{\label{fig4}
Differential conductance correction $dI/dV$ 
per defect for bias $V$ near the renormalized level 
spacing $\Delta(D=0)$, $V>T_K  \approx 10^{-4}D_0$. 
$J_{z}/2= J_{\perp}=0.016$, $g _{01}^{(j)}= g_{10}^{(j)} = 0.6 J_{\perp}$,
$\Delta_0= 3\, T_K$, $\Delta (D=0)= -1.82\, T_K$. 
Right inset: $dI/dV$ for a larger bias range. 
The dashed line is a sketch of the 2CK-induced $\sqrt{|V|}$--ZBA of
width $T_K$, with the numerically determined $T_K$, as indicated. 
Left inset: Experimental $dI/dV$ curve taken from 
Ref.~\cite{ralph92} for comparison. 
}
\end{figure}

{\it Nonequilibrium differential conductance. ---}
Finally we compute the current through an 
ultrasmall metallic point contact in the presence of 2CK defects.
We assume that a conductance channel is blocked when a
2CK scattering from that channel occurs. The current correction
due to a single 2CK defect is then 
$\delta I(V) = - \frac{e}{\hbar} \left< [(\rho _+ -\rho_- ), H ] \right>$,
with $\rho_{\alpha}={\sum_{{\bf k}\sigma m}}' 
c^{\alpha\ \ \dagger}_{{\bf k}\sigma m}
c^{\alpha\ \ \phantom{\dagger}}_{{\bf k}\sigma m}$. 
We computed this expression for $T=0$, as outlined in Ref.~\cite{rosch03}, 
by evaluating the 2nd order Keldysh diagram shown in Fig.~\ref{fig1}b)  
\cite{paaske04a}, using the renormalized  
coupling functions obtained from Eq.~(\ref{RG}), and obtained the 
voltage dependent nonequilibrium conductance $dI/dV$. 
Details of the lengthy calculation will be presented 
elsewhere.
The results are displayed in Fig.~\ref{fig4}, showing   
conductance spikes much narrower than $T_K$ at a bias $V$ corresponding 
to the renormalized level spacing for that bias, $V=\pm|\Delta |$.
These spikes, with a steep step on the low-$V$ side
and a wider (logarithmic) decay on the high-$V$ side, are analogous to
the conductance peaks of a Kondo quantum dot in a magnetic field
\cite{rosch03}. They are due to transitions between the 
(renormalized) ground- and excited-state defect levels, enhanced by 
Kondo scattering. Despite their complicated shape arising from 
a combination of many terms, their overall narrow 
width can be understood from the fact that the model (\ref{hamiltonian}) 
contains at least one more low-energy scale in addition to $T_K$:
While $T_K$ is to leading log order determined by fluctuations 
between the {\it degenerate} defect levels ($m=\pm 1$), i.e. by $J_{\perp}$, 
$T_K\simeq D_0 \exp [-1/4J_{\perp}]$, the conductance has in 3rd  
order [cf. Fig.~\ref{fig1}c)] a logarithmic divergence at $V=\pm |\Delta|$, which is governed
by the couplings $g_{10}^{(1)}=g_{-10}^{(-1)}$, \ 
$dI/dV\simeq 4g_{10}^{(1)}\ \ln [\Theta(|V|-|\Delta|)(|V|-|\Delta|)/D_0]$.
The width of this peak, although additionally cut off by the nonequilibrium 
rates $\Gamma _{m}$, may thus be estimated as 
$T_K^{\star}\simeq D_0 \exp [-1/4g_{10}^{(1)}]$, 
i.e. for $g_{10}^{(1)} < J_{\perp}$ it is 
exponentially smaller than $T_K$ \cite{footnote}.

In conclusion, we have shown 
that a three level defect with partially broken SU(3) symmetry and
degenerate excited states in a metal 
has a sizable, stable 2CK fixed point phase and that this 
model can describe conductance spikes exponentially more narrow
than $T_K$. 
Hence, the model suggests a unified microscopic explanation 
for the essential features of the conductance anomalies observed by Ralph 
and Buhrman \cite{ralph92}. The reduction of the ZBA and the splitting 
of the spikes in a magnetic field $B$ \cite{ralph92}
can be explained  
by a coupling of $B$ to the defect angular momentum
and the resulting Zeeman splitting of the levels $m=\pm 1$. 
This will be discussed elsewhere.
The size of the anomalies $\sim$ 100 $e^2/h$
might be due to a corresponding number of 2CK defects in the
contact. 
The very sharp experimental distribution of $T_K$ \cite{ralph92}  
may occur, because the rotational defects form at equivalent 
interstitial sites in a (nearly perfect) Cu lattice or because 
only those defects with sufficiently high $T_K$ are 
observable \cite{kolf07}. ---
We wish to thank J. von Delft, A. Rosch, and especially A. Zawadowski for
numerous useful discussions. This work was supported by 
DFG through SFB 608 and grant No. KR1726/1.


\begin{thebibliography}{10}
\bibitem{footnote}
In case $\Gamma _m (V) > T_K^{\star}$, the peak width is controlled 
by $\Gamma _m (V)$ and may be significantly larger than $T_K$.
\end{thebibliography}
\end{document}